\documentclass[twocolumn,floats,floatfix,superscriptaddress,aps,pra]{revtex4-2}
\usepackage{subfigure}
\usepackage{psfrag,graphicx}
\usepackage{bm,color}
\usepackage{latexsym}
\usepackage{epstopdf}
\usepackage[english]{babel}
\usepackage{amsmath,amssymb,amsfonts}
\usepackage{appendix}
\usepackage{bbold}
\usepackage[dvipsnames]{xcolor}

\definecolor{mygrey}{gray}{0.35}
\definecolor{myblue}{rgb}{0.2,0.2,0.8}
\definecolor{myzard}{cmyk}{0,0,0.05,0}
\definecolor{mywhite}{rgb}{1,1,1}
\definecolor{mywhite}{rgb}{1,1,1}
\definecolor{myred}{rgb}{1,0.,0.3}

\def\be{ \begin{equation}}
\def\ee{ \end{equation}}
\def\bse{  \begin{subequations}}
\def\ese{  \end{subequations}}
\def\bea#1\ea{\begin{align}#1\end{align}}

\def\bi{\begin{itemize}}
\def\ei{\end{itemize}}

\def\bt{\begin{tabular}}
\def\et{\end{tabular}}

\def\ket#1{\vert #1 \rangle}

\def\half{\tfrac12}

\def\3half{\tfrac32}

\def\Black{}

\def\t{  (t)}

\def\from{\leftarrow}
\def\to{\rightarrow}

\def\A{\mathbf{A}}
\def\B{\mathbf{B}}
\def\U{\mathbf{U}}
\def\C{\mathbf{C}}
\def\H{\mathbf{H}}
\def\S{\mathbf{S}}
\def\C{\mathbf{C}}
\def\D{\mathbf{D}}

\def\V{\mathbf{V}}

\def\R{\mathbf{R}}

\def\F{\mathbf{F}}

\def\R{\mathbf{R}}

\begin{document}
\author{K. N. Zlatanov}
\affiliation{Department of Physics, Sofia University, James Bourchier 5 blvd., 1164 Sofia, Bulgaria}
\affiliation{Institute of Solid State Physics, Bulgarian Academy of Sciences, Tsarigradsko chauss\'ee 72, 1784 Sofia, Bulgaria}
\author{A. A. Rangelov}
\affiliation{Department of Physics, Sofia University, James Bourchier 5 blvd., 1164 Sofia, Bulgaria}
\author{N. V. Vitanov}
\affiliation{Department of Physics, Sofia University, James Bourchier 5 blvd., 1164 Sofia, Bulgaria}
\title{Extension of the Morris-Shore transformation to arbitrary time-dependent driving fields}
\date{\today }

\begin{abstract}
The treatment of time-dependent dynamics of quantum systems involving multiple states poses considerable technical challenges.
One of the most efficient approaches in treating such systems is the Morris-Shore  (MS) transformation which decomposes the multistate dynamics to a set of independent systems of simpler interaction pattern and uncoupled spectator states. The standard MS transformation imposes restrictions on the time dependence of the external fields addressing the states, as it requires that both Rabi frequencies have the same time profile. 
In this work we treat the case of the time-dependent MS transformation, which opens prospects for a variety of physically interesting processes wherein the fields may have different time dependences.
We explore the adiabatic and the double-adiabatic limit, in which we demonstrate population transfer between the MS states that results in population transfer from one set of states onto another. 
We demonstrate the generation of superposition states between the MS states by the techniques of half adiabatic passage and fractional stimulated Raman adiabatic passage, which translate to superpositions of all the states of the involved levels.
\end{abstract}

\maketitle


\section{Introduction}

Coherent control of  quantum systems is of paramount importance in contemporary quantum physics \cite{Allen,Shore} and its most prominent part, quantum information \cite{Nielsen2000}. 
The most notable effects come from systems that have been steered into a coherent superposition state. 
In the most common scenario the superposition is induced in a two-state system, or qubit. 
Two-state systems are very well understood and can be controlled with high accuracy by a variety of techniques. 
For example, resonant excitation can generate superposition states by applying a $\pi/2$ pulse. 
When robustness is crucial adiabatic excitation proves to be very useful, since the superposition state depends on the initial and final values of the laser pulses. 
This prevents imperfections in the laser field to alternate the final state. 
In a  two-state system rapid adiabatic passage   (RAP) techniques can be used in various ways to generate superpositions \cite{Zlatanov2017,Zlatanov2020b}. 
Similar control can be exerted by composite pulses and inverse engineering  techniques   (e.g. optimal control, ``shortcuts to adiabaticity'', etc.).
Three-state systems are also well studied and they add more control complexity but offer in return far more flexibility and intriguing new features and opportunities, e.g. the famous technique of stimulated Raman adiabatic passage   (STIRAP) \cite{Vitanov2017}, which can also generate superposition states in a configuration known as fractional STIRAP \cite{Vitanov1999}.

Quantum systems with more than three states, as encountered in real quantum systems, are far more difficult to treat analytically in the general case due to their mathematical complexity. 
In some special cases, though, they can be reduced to simpler two- and three-state systems which makes their analytic study feasible.
Such systems are enjoying rapidly increasing interest in qudit   ($d$-state quantum system) quantum information \cite{Wang2020}. 

The general approach to systems with multiple states is to search for a change of basis which transforms the original states to a superposition of states and reduces the number of interactions.
One of the most efficient approaches is the Morris-Shore   (MS) transformation \cite{Morris83,Shore2013}.
It is applicable to multistate systems, in which any state in one set may be coupled to any state in another set, although not all states have to be intercoupled. 
The MS transformation decomposes the dynamics of such multistate systems to a set of independent systems and uncoupled single   (or dark) states. 
The original application of the MS transformation is in problems having their states distributed over two sets of states. This has been extended to the situation in which the states are distributed among $N$ sets \cite{Rangelov2007} with a similar decomposition of the dynamics into a set of independent $N$-state systems of simpler interaction pattern.

The general requirements for the existence of a MS transformation can be classified in relation to the properties of the system and properties of the excitation fields. The system is required to be degenerate, although this has been relaxed recently \cite{Zlatanov2020a}, and also cross couplings between states of the same set are forbidden. The requirements for the laser fields dictate the same time dependence of the Rabi frequencies and the same time dependence of the detunings.  
This condition impedes the application of the MS transformation to various interesting processes wherein the fields have different time dependences, e.g. as in RAP or STIRAP and their extensions \cite{Vitanov2017}, e.g. tripod-STIRAP \cite{Unanyan1998}.

The present paper extends the MS transformation to such situations, in which all couplings can have different time dependences.
We treat adiabatic excitation of the system, which grants us a large system reduction while allowing the realisation of adiabatic techniques like RAP and STIRAP. 
We demonstrate how under the proper conditions we can transfer a superposition state composed of the states in one level onto another. 
We also show how a superposition state from all of the states in the interacting levels can be formed.
This method can be useful in the rapidly developing field of qudit quantum information \cite{Brukner,Vaziri}, as qudits --- $d$-dimensional quantum systems with $d>2$ --- offer much larger Hilbert space and some other advantages over qubits   ($d=2$). 

This paper is organized as follows. 
In Section~\ref{sec:two} we introduce the problem of multistate excitation, we introduce the time-dependent MS transformation, and discuss the evolution of the system under adiabatic conditions. 
In Section~\ref{sec:three} we demonstrate how the time-dependent MS transformation can generate various superposition states in three common and different configurations, namely an N-pod linkage, as well as W and X linkages. 
Finally, we conclude in Section~\ref{sec:four}.

\section{Dynamics in the Morris-Shore basis}\label{sec:two}


The excitation dynamics of a system under the influence of laser excitation is described by the Schr\"{o}dinger equation, which in the rotating-wave approximation   (RWA) \cite{Allen,Shore} takes the form of coupled ordinary differential equations for time-dependent complex-valued probability amplitudes $\mathbf{c}   (t) = [c_1  (t), c_2  (t), \ldots, c_{n}  (t)]^T$,
\begin{equation}
i\frac{d}{dt}\mathbf{c}   (t) = \mathbf{H}  (t) \mathbf{c}   (t) ,
\end{equation}
where we have set $\hbar=1.$
Systems with states distributed in two and three levels have Hamiltonians in the respective forms%
\bse\label{H}
\begin{equation}
\mathbf{H}=\frac{1}{2}%
\begin{bmatrix}
\mathbf{0} & \mathbf{V}  (t)\\
\mathbf{V}^{\dag }  (t) & 2\mathbf{\Delta }  (t)%
\end{bmatrix}%
,  \label{H_2lvl}
\end{equation}\\
for two levels and 
\begin{equation}
\mathbf{H}=\frac{1}{2}\left[
\begin{array}{ccc}
 \mathbf{0} & \V_p  (t) & \mathbf{0} \\
 \V _p^{\dag}  (t) & 2\mathbf{\Delta}  (t) & \V_s  (t) \\
 \mathbf{0} & \V_s^{\dag}  (t) & \mathbf{0} \\
\end{array}
\right],  \label{H_3lvl}
\end{equation}
\ese
for three levels, where we have allowed only the middle set of states to be off resonance.
The interaction matrices $\mathbf{V}  (t)$ are $g\times e$-dimensional and are composed of the couplings between the $g$ and $e$ sets of states. 
The detuning matrix in Eqs.~\eqref{H} is diagonal,
\be
\mathbf{\Delta}  (t)=\Delta  (t) \mathbf{1}.
\label{delta-mat}
\ee
Its   (generally time-dependent) diagonal elements are defined as the difference of the transition frequency $\omega_0$ of the system and the frequency $\omega$ of the coupling field,
\be
\Delta  (t) = \omega_0-\omega,
\ee
When multiple states are involved the dynamics of the system becomes very complicated and usually analytical solutions can not be derived. 
In such cases, switching to the MS basis can drastically simplify the problem by reducing the system to independent sub-systems of smaller dimension.

\begin{figure}[tb]
\bt{cc}
\includegraphics[width=0.85\columnwidth]{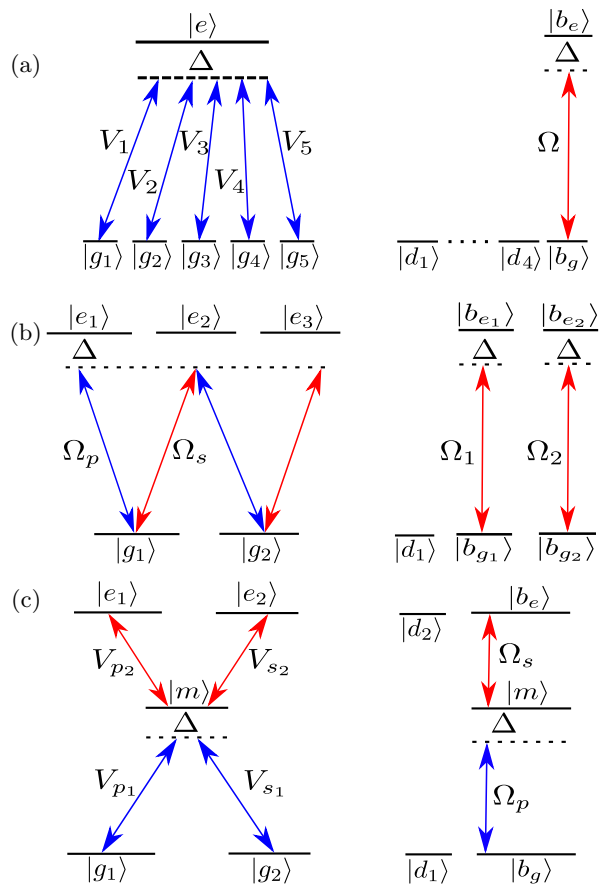}\llap{
  \parbox[b]{6.0in}{  (a)\\\rule{0ex}{4.21in}
  }}\llap{
  \parbox[b]{6.0in}{  (b)\\\rule{0ex}{2.81in}\llap{
  \parbox[b]{0.075in}{  (c)\\\rule{0ex}{1.41in}
  }}
  }}
\et
\caption{  (Color online)
Multilevel systems in different configurations. Left: Systems in   (a) Five-pod,   (b) W and   (c) X linkage. Right: The respective Morris-Shore transformed systems.
}\label{fig:1}
\end{figure}

\subsubsection{The static Morris-Shore transformation}
The aim of the MS transformation is to find a transformation matrix $\S$ such that the state vector
\be
\C^{MS}=\mathbf{S}\C=\left[\D,\B\right]^T,
\ee
will be partitioned into $\D$ dark and $\B$ bright MS states that can interact as shown in Fig.~\ref{fig:1}.
The MS Hamiltonian can be arranged into the block diagonal form 
\be
\H^{MS}=\left[
\begin{array}{cccc}
 \H^d & 0 & \hdots & 0 \\
  0 & \H^{b}_{1} & \hdots & 0 \\
   \vdots &  \vdots  & \ddots   & \vdots\\
 	0 & 0 & \hdots & \H^{b}_{n} \\
\end{array}
\right].
\ee
Here $\H^d$ is diagonal and governs the evolution of the dark states, while the $\H^{b}_n$ bright Hamiltonians drive the evolution of the bright states. 

In order to find the map $\S$ between the MS basis and the original basis we look for a matrix that has a diagonal block structure which for two levels has the form
\bse
\be
\S_{2lvl}=\left[
\begin{array}{cc}
\mathbf{A} & \mathbf{0} \\
\mathbf{0} & \mathbf{B}%
\end{array}%
\right] ,  \label{R_2lvl}
\ee%
and 
\be
\S_{3lvl}=\left[
\begin{array}{ccc}
 \A & \mathbf{0} & \mathbf{0} \\
 \mathbf{0} & \B & \mathbf{0} \\
 \mathbf{0} & \mathbf{0} & \F \\
\end{array}
\right]\label{R_3lvl}
\ee
\ese
for three levels.
The matrices $\A,\B$ and $\F$ have to be such that the detuning in the bright Hamiltonians is the same as in the original basis.
In the two-state case $\S$ has to diagonalize the matrices of the original Hamiltonian such that the new coupling matrix 
\be
\mathbf{\Omega}=\A\V\B^{\dagger},
\ee
is now diagonal.
Alternatively in the three-state case the sub-matrices of $\S$ have to diagonalize
\bse
\be
\mathbf{\Omega_p}=\A\V_p\B^{\dagger},
\ee
\be
\mathbf{\Omega_s}=\B\V_s\F^{\dagger}.
\ee
\ese
This diagonalization is only possible if 
\be
\Big[\V_p^{\dagger}\V_p,\V_s\V_s^{\dagger} \Big]=0
\ee
holds.

\Black

Depending on the linkage pattern, the bright Hamiltonians can act on two or three MS states and take the respective forms
\bse
\be
\H^{b}_{2lvl}=\frac{1}{2}%
\left[
\begin{array}{cc}
0 & \Omega  (t) \\
\Omega  (t) & 2\Delta  (t)%
\end{array}%
\right],\label{HMS2lvl}
\ee\\
and
\be
\H^{b}_{3lvl}=\frac{1}{2}\left[
\begin{array}{ccc}
 0 & \Omega _p  (t) & 0 \\
 \Omega _p  (t) & \Delta   (t) & \Omega _s  (t) \\
 0 & \Omega _s  (t) & 0 \\
\end{array}
\right].\label{HMS3lvl}
\ee
\ese
The new interaction elements $\Omega  (t)$ on the other hand are a function of the elements of $\V  (t)$ and depend on the linkage pattern. 

\subsubsection{The time-dependent Morris-Shore transformation}
The derivation of the static MS transformation has two requirements. First the system has to be degenerate, which can be relaxed
as shown in \cite{Zlatanov2020a} and second, the time dependence of the Rabi frequencies has to be shared with the detuning.
In order to generalize the earlier result we allow each Rabi frequency as well as the detuning to have
a different time dependence. 
The MS transformation matrix $\S  (t) $ between the original basis and MS basis will then be time dependent.
The  MS Hamiltonian then reads
\begin{equation}
\H^{MS}  (t)=\S  (t)\H  (t)\S^{\dagger }  (t)-i\S  (t)\frac{d}{dt}\S^{\dagger }  (t).\label{Time MS}
\end{equation}%
 The second term on the right hand side imitates non-adiabatic couplings amongst the dark states, and amongst the dark states
 and the bright states. These couplings undermine the benefits of the MS transformation since an overall simplification of the system has not occur.
In order to overcome this problem we can impose specific limitations on the laser pulses such that
\be
\S  (t)\frac{d}{dt}\S^{\dagger }  (t)\approx 0.\label{MSadiabatic}
\ee
Exciting the system adiabatically can achieve such elimination.  The restrictions Eq.  (\ref{MSadiabatic}) imposes depend on the specific linkage pattern of the system, but in general translate to a large combined pulse area. 
An alternative way to eliminate the extra term on the right hand side of Eq.  (\ref{Time MS}) is to have the same time dependence for every coupling. Then although the detuning and the couplings have different time dependence the additional non-adiabatic term vanishes and Eq.  (\ref{MSadiabatic}) is exact. This way we reduce the system to the static case only this time we are much less restricted. 
\subsubsection{Double-adiabatic Morris-Shore evolution}
In order to utilize the time dependent MS transformation it is worthwhile to transfer the MS Hamiltonian to the adiabatic basis. Since we already imposed adiabatic evolution  to comply with Eq.  (\ref{MSadiabatic}), this second transformation we will call double-adiabatic   (DA), in order to avoid confusion with the term "superadiabatic".
In the double-adiabatic basis the MS Hamiltonians transform in a similar way to the time dependent MS Hamiltonian as
\be
\H^{DA}  (t)=\U  (t)\H^{MS}  (t)\U^{\dagger }  (t)-i\U  (t)\frac{d}{dt}\U^{\dagger }  (t).\label{HSAextra}
\ee
The structure of $\U  (t)$ depends on the number of levels and dark states involved. It is composed of an identity operator acting on the dark MS Hamiltonian
and a number of rotation matrices that act on the $k$ bright Hamiltonians. In the two-state case the rotations have the form
\be
\R  (\alpha_k)=\left[
\begin{array}{cc}
\cos  (\alpha_k) & \sin  (\alpha_k) \\
-\sin  (\alpha_k) & \cos  (\alpha_k)%
\end{array}%
\right],\label{Rot2lvl}
\ee
where the $k$-th double-adiabatic angle reads
\be
\alpha_k=\frac{1}{2} \tan ^{-1}\left  (\frac{\Omega_k  (t)}{\Delta   (t)}\right).\label{angle2lvl}
\ee
When the bright Hamiltonian is the one of Eq.  (\ref{HMS3lvl}), the $k$-th three-state rotation becomes
\be
\R  (\theta_k,\phi_k)=\left[
\begin{array}{ccc}
 \sin   (\theta_k) \sin   (\phi_k ) & \cos   (\phi_k ) & \cos   (\theta_k ) \sin   (\phi_k ) \\
 \cos   (\theta_k ) & 0 & -\sin   (\theta_k) \\
 \cos   (\phi_k ) \sin   (\theta_k) & -\sin   (\phi_k ) & \cos   (\theta_k) \cos   (\phi_k ) \\
\end{array}
\right],\label{Rot3lvl}
\ee
where the double-adiabatic angles read
\bse\label{angle3lvl}
\be
\theta_k=\tan ^{-1}\left  (\frac{\Omega _{p,k}  (t)}{\Omega _{s,k}  (t)}\right),
\ee
\be
\phi_k=\frac{1}{2} \tan ^{-1}\left  (\frac{\sqrt{\Omega _{p,k}  (t){}^2+\Omega _{s,k}  (t){}^2}}{\Delta   (t)}\right).
\ee
\ese
In order to eliminate the coupling in the double-adiabatic basis we have to impose new conditions on the laser pulses.
In general the double-adiabatic condition reads
\be
\left\vert \left\langle \Phi_{j}\left  ( t\right) |\frac{d}{dt}\Phi_{k}%
\left  ( t\right) \right\rangle \right\vert \ll \epsilon_k,
\label{MSSupAd}
\ee
where $\Phi_j$ and $\Phi_k$ are the DA states composed of the MS states and $\epsilon_k$ is the energy of the $k$-th double-adiabatic state.
In practice the condition of Ineq.  (\ref{MSSupAd}) can be simplified when we transform the bright Hamiltonians. For the two-level MS systems of Eq.  (\ref{HMS2lvl}) the double-adiabatic condition reads
\bse
\be
\left|\Delta   (t)\frac{d}{dt}\Omega  (t)-\frac{d}{dt}\Delta  (t) \Omega  (t)\right| \ll \Big[\Omega  (t){}^2+\Delta  (t)^2\Big]^{1/2}.\label{adcond2lvl}
\ee
If we instead evolve the MS Hamiltonian of Eq.  (\ref{HMS3lvl}) double-adiabatically, 
\be
\left| \Omega _p  (t) \frac{d}{dt}\Omega _s  (t)-\Omega _s  (t) \frac{d}{dt}\Omega _p  (t) \right| \ll \Omega _p  (t){}^2+\Omega _s  (t){}^2\label{adcond3lvl}
\ee
has to be satisfied.
\ese
Eliminating the off-diagonal couplings in the double-adiabatic basis we reduce Eq.  (\ref{HSAextra}) to
\be
\H^{DA}  (t)=\U  (t)\H^{MS}  (t)\U^{\dagger }  (t),\label{HSA}
\ee
which is a diagonal matrix.
Now the advantage of this two step transformation to the MS and then to the double-adiabatic basis becomes clear. We can demonstrate the well known adiabatic techniques
RAP and STIRAP. This turns out to be very flexible and above all robust way to generate a superposition of states since the double-adiabatic states are composed of MS states. For example complete population transfer   (CPT) will map a superposition of one set of states to another. Alternatively if we stop the excitation half-way, which means to generate a superposition between the MS states, we end up with a superposition between all the states in the interacting levels. 
\Black




\section{Examples}\label{sec:three}

In this section we demonstrate the approach given above by applying it to a few different types of systems, namely a N-pod linkage, a two-level W and three-level X linkage.
The time-dependent MS transformation allows us a great freedom in the time profiles of the laser pulses and their chirping, however it also complicates the dynamics drastically. 
In order to keep the examples simple we will only treat excitations of two or three different profiles, either for $V  (t)$ and $\Delta  (t)$ or the pump $V_p  (t)$, the Stokes $V_s  (t)$ and the detuning. 

\subsection{N-pod linkage}
The general form of a Hamiltonian whose states in the original basis are linked through a common state, shown in Fig.~\ref{fig:1}   (a),  reads 
\be
\H=\half\left[
\begin{array}{cccc}
 0 & \hdots & \hdots & V_1  (t) \\
  \vdots & 0& \hdots & \vdots \\
   \vdots &  \hdots  & \ddots   & V_n  (t)\\
 	V_1  (t) & \hdots & V_n  (t) & 2\Delta  (t) \\
\end{array}
\right].
\ee
If we excite the system such that Eq.  (\ref{MSadiabatic}) holds, then the MS Hamiltonian reduces it to a dark Hamiltonian and a single two-state bright Hamiltonian 
\be
\H^{MS}  (t)=\frac{1}{2}\left[
\begin{array}{cc}
\H^d & 0 \\
0 & \H^b%
\end{array}%
\right] .  \label{NpodH}
\ee%

In general, if we initialise the system in a bright state, then we are only concerned with the evolution of the two-state system. 
The N-pod system has the unique property that allows us to initialize the system in the ground bright state without creating an initial superposition between the states of the original basis. That means we can have the population in, say $\ket{g_1}$ and still steer the evolution by $\H^b$ without populating any of the dark states. This becomes apparent from the explicit form of the dark states, which in the basis discussed in \cite{Kyoseva2006} read, 
\bea\label{dark states Npod}
& \ket{d_{1}}  =\frac{\Big[ V_{2} \t ,-V_{1} \t
,0,0, \ldots ,0\Big] ^{T}}{\Omega_{2} \t  }
,
\\
& \ket{d_{2}}  =\frac{\Big[ V_{1} \t V_{3} \t , V_{2} \t V_{3} \t ,-\Omega_{2}^{2} \t ,0, \ldots ,0\Big] ^{T}}{\Omega_{2} \t \Omega_{3} \t }
,
\\
& \ket{d_{3}}   =\frac{ \Big[ V_{1} \t V_{4} \t ,V_{2} \t V_{4} \t ,
 V_{3} \t V_{4} \t ,-\Omega_{3}^{2} \t ,0,
\ldots ,0\Big] ^{T}}{\Omega_{3} \t \Omega_{4} \t },
 \\
& \vdots  
\notag 
\\
& \ket{d_{N-1}}   =\frac{ \Big[ V_{1}\t  V_{n}\t ,\ldots ,  V_{n-1}\t  V_{n}\t ,-\Omega_{n-1}^{2}  ,0\Big] ^{T}}{\Omega_{n-1} \t \Omega_{n} \t },
\ea
where
\begin{equation}
\Omega_{n} \t  = \sqrt{\sum_{k=1}^{n}V_{k}^{2} \t }, \quad   n=2,3,\ldots ,N.
\end{equation}
The ground bright state also reads
\be\label{bright state Npod}
 \ket{b_g} =\frac{\Big[ V_{1} \t ,V _{2} \t ,\ldots ,V_{n} \t,0\Big] ^{T}}{%
\Omega_{n} \t }.
\ee

Note that each component of the dark states and the ground bright state are dependent on the Rabi frequencies. This can be exploited to drive the system in $\ket{b_g}$ if we let the pulse that couples $\ket{g_1}$ and $\ket{e}$ start before all others. 
This pulse arrangement, does not allow the dark states to be populated and provides a way to generate a superposition between all states in the ground level by an excitation that achieves complete population return  (CPR). In a five-pod system such excitation will eventually result in a final state that reads,
\be
\ket{b_g}=\frac{\displaystyle\sum_{i=1}^5V_i\t\ket{g_i}}{\Omega_5\t}.
\ee
In a similar way to the initialisation in the ground bright state, we can control the final superposition by letting the pulses coupling undesired components to vanish before the rest, while the rest coincide in the final moment. For example if $V_1  (t),V_2\t$ and $V_4  (t)$ end before the rest, we can generate
\be
\ket{b_g}=\frac{\ket{g_3}+\ket{g_5}}{\sqrt{2}}.\label{Npod-2states superposition}
\ee 
\begin{figure}[tb]
\bt{cc}
\qquad \quad \includegraphics[width=0.73\columnwidth]{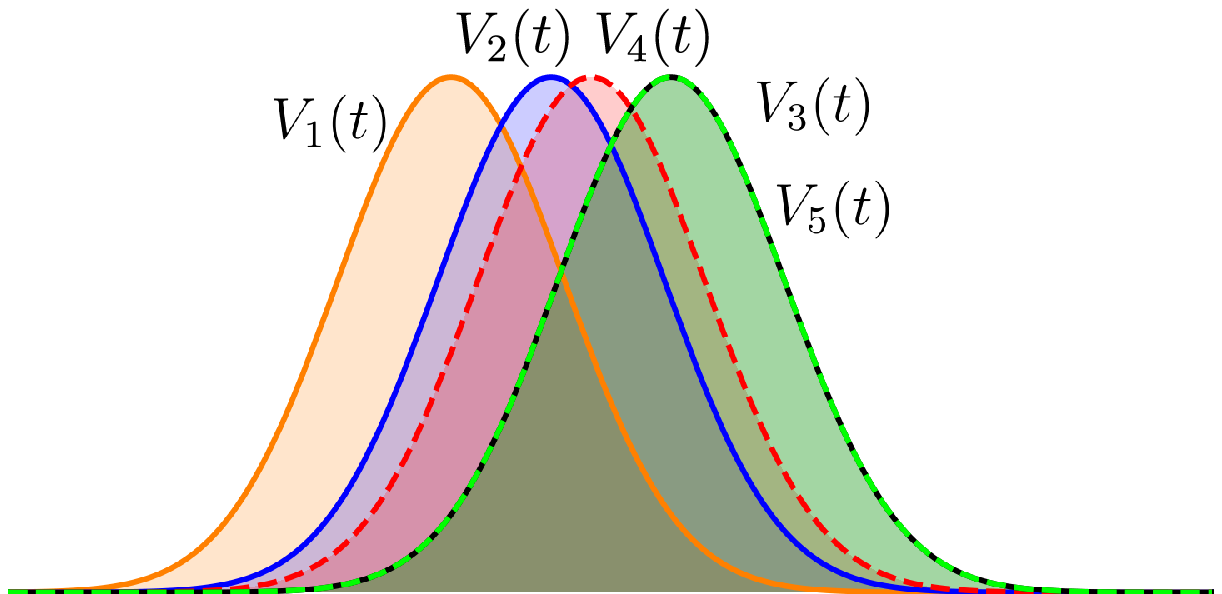}\llap{
  \parbox[b]{4.5in}{  (a)\\\rule{0ex}{1.01in}
  }}\vspace{.2cm}\\
  \qquad\quad \includegraphics[width=0.73\columnwidth]{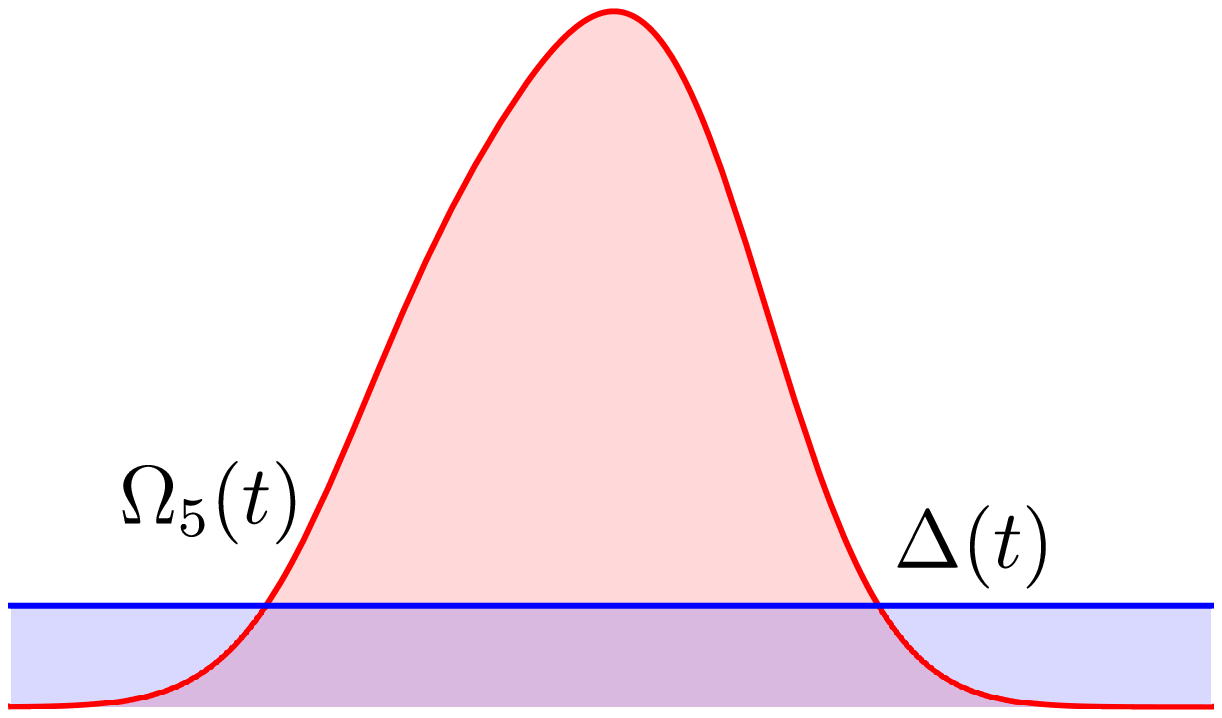}\llap{
  \parbox[b]{4.5in}{  (b)\\\rule{0ex}{1.01in}
  }}\vspace{.2cm}\\
  \includegraphics[width=0.86\columnwidth]{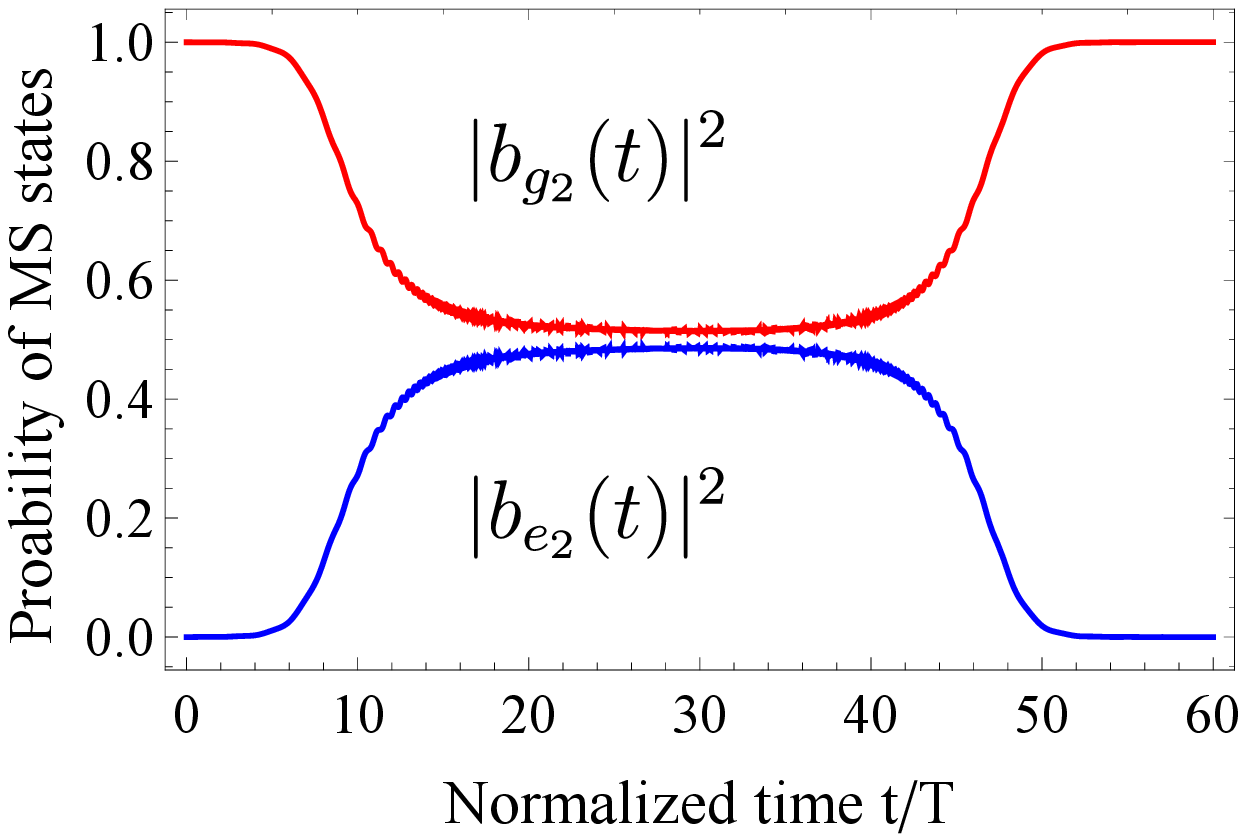}\llap{
  \parbox[b]{4.6in}{  (c)\\\rule{0ex}{1.601in}
  }}  
\et
\caption{  (Color online)
Complete population return  (CPR) for the ground bright state of a five pod system.
  (a) The original couplings initialize the system in $\ket{b_g},$ by having $V_{1}  (t)$ start before the rest of the interactions. Eventually $V_{1}  (t), V_2\t$ and $V_4  (t)$ are switched off, while $V_3\t$ and $V_5\t$ coincide, so the final superposition is the one of Eq.  (\ref{Npod-2states superposition}). The   (b) frame shows the MS Rabi frequency and the detuning.  
The   (c) frame demonstrates the CPR of the bright states. The simulation is carried with constant detuning $\Delta=4.33 T^{-1},$  and Gaussian pulses of the form $V_{i}\t=V_i^0\exp  (-  (t-\tau_i)^2/T^2)$ where all of the couplings share the same magnitude $V_i^0=60 T^{-1}$ and the pulses are shifted by $\tau_{1}=22 T^{-1},\tau_{2}=27 T^{-1}$ and $\tau_{3}=29T^{-1}, \tau_{4}=\tau_5=33 T^{-1}.$  
}\label{fig:2}
\end{figure}
The illustration of superposition states is often a nontrivial task. At best a depiction of a Bloch vector on a sphere can be given or alternatively a plot of the populations is drawn. Both approaches are not convenient when more than two states are involved. In the case of the Bloch vector, one has to draw too many coherences, and population differences since its components grow with $N^2-1,$ where $N$ is the number of states. The problem with plotting the populations streams from the fact that they are the modulus squared of the probability amplitudes and when multiple states are involved it becomes hard for the reader to factorize back the superposition states. The MS states on the other hand can still be pictured as a population plot, where CPR or CPT mean that we have populated a specific superposition. In Fig.~\ref{fig:2} we demonstrate the generation of the superposition of Eq.  (\ref{Npod-2states superposition}) with CPR. The simplicity of the N-pod system allows us to generate different superpositions by only imposing the adiabatic condition of Eq.(\ref{MSadiabatic}). For more complex systems the adiabatic condition has to be extended over the detuning as we show in the next two sections.

\subsection{W linkage in two levels}
In the previous example we looked at an N-pod system, which simplified the system by leaving only a single pair of bright MS states. In this example we investigate the W-linkage in two levels depicted in Fig~\ref{fig:1}   (b). The Hamiltonian reads
\be
\H_W=\half\left[
\begin{array}{ccccc}
 0 & 0 & V_p  (t) & V_s  (t) & 0 \\
 0 & 0 & 0 & V_p  (t) & V_s  (t) \\
 V_p  (t) & 0 & 2 \Delta   (t) & 0 & 0 \\
 V_s  (t) & V_p  (t) & 0 & 2 \Delta   (t) & 0 \\
 0 & V_s  (t) & 0 & 0 & 2 \Delta   (t) \\
\end{array}
\right].
\ee
Since we excite the system adiabatically the actual pulse shape is not important as long as the adiabatic condition of Eq.  (\ref{MSadiabatic}) is satisfied.
In this example 
\be
\frac{V_s  (t) V_p'  (t)+V_p  (t) V_s'  (t)}{\sqrt{V_p  (t){}^2+V_p  (t) V_s  (t)+V_s  (t){}^2}}\approx0,
\ee
has to hold.
However for the sake of simplicity we assume that $V_i  (t)$, $i=s,p,$ is real and positive, which is well justified for say Gaussian pulses. 

Switching to the MS basis, under adiabatic evolution then reads,
\be
\H^{MS}_W=\half\left[
\begin{array}{ccc}
 \H^b_1 & 0 & 0 \\
 0 & 2\Delta   (t) & 0 \\
 0 & 0 & \H^b_2 \\
\end{array}
\right]\label{W-HMS},
\ee
where the bright Hamiltonians have the form of Eq.  (\ref{HMS2lvl}) and the new coupling elements are given as
\bse
\be
\Omega_1  (t)= \sqrt{V_p  (t){}^2-V_p  (t) V_s  (t)+V_s  (t){}^2},
\ee
and
\be
\Omega_2  (t)=\sqrt{V_p  (t){}^2+V_p  (t) V_s  (t)+V_s  (t){}^2} .
\ee
\ese
The MS state vector, expressed in the original basis states, reads 
\be
\C^{MS}_W=\frac{1}{\sqrt{2}}\left[
\begin{array}{c}
 \ket{g_2}-\ket{g_1}\\
 \frac{\left[\ket{e_2}-\ket{e_1}\right] V_p  (t)+\left[\ket{e_3}-\ket{e_2}\right]
   V_s  (t)}{\Omega_1  (t)} \\
\sqrt{2} \frac{V_s  (t) \left[\ket{e_1} V_s  (t)-\ket{e_2} V_p  (t)\right]+\ket{e_3} V_p  (t){}^2}{\Omega_1  (t)\Omega_2  (t)} \\
 \ket{g_2}+\ket{g_1} \\
 \frac{\left[\ket{e_1}+\ket{e_2}\right] V_p  (t)+\left[\ket{e_2}+\ket{e_3}\right]
   V_s  (t)}{\Omega_2  (t)} \\
\end{array}
\right].\label{W-MS states}
\ee
The next step is to switch to the double-adiabatic basis. The DA Hamiltonian, which is a diagonal matrix, is not that important since it only contributes with a global phase factor to the final state. Instead we turn our attention to the evolution of the DA state vector, which in terms of the MS states reads,
\be
\C^{DA}  (t)=\U  (t)\C^{MS}=\left[
\begin{array}{ccc}
 \R\left  (\alpha _1\right) & 0 & 0 \\
 0 & 1 & 0 \\
 0 & 0 & \R\left  (\alpha _2\right) \\
\end{array}
\right]\left[\begin{array}{c}
 \ket{\B_1} \\
 \ket{d} \\
 \ket{\B_{2}}\\
\end{array}\right],\label{W-DA states}
\ee
where the $\ket{\B_i}=[\ket{b_{g_i}},\ket{b_{e_i}}]$ vectors are composed of the interacting bright states and the rotation matrices have the form of Eq.  (\ref{Rot2lvl}).
We have assumed that the double-adiabatic condition of Eq.  (\ref{adcond2lvl}) holds for $\Omega_1  (t),\Omega_2  (t)$ and $\Delta  (t).$  
The final MS state, and consequently the superposition of states in the original basis are now determined according to the initialisation of the system and the values of the double-adiabatic angles $\alpha_i,$ given by Eq.  (\ref{angle2lvl}).\\
We have two options for initialising the system.
First, to initialise in any of the states of the original basis, which proves to be unreliable. For example the excited states will populate the dark state as well the excited bright states. Alternatively we can initialise the system in any of the  ground states and attempt to use the same strategy as in the N-pod system. Letting the coupling field to start prematurely, we will activate both blocks of Eq.  (\ref{W-HMS}), and therefore a controllable procedure for generating a specific superposition becomes dubious.
Second, we can initialise the system directly in a bright state. This way we ensure that only a single pair of the MS states will participate in the interaction and we can control the superposition with the double-adiabatic angle. We can have the following superpositions:

\emph{Superposition transfer by RAP.} If we initialise the system in state $\ket{b_{g_2}}=\frac{\ket{g_2}+\ket{g_1}}{\sqrt{2}}$ only the lower block of Eq.  (\ref{W-HMS}) steers the evolution of the system. Consequently in the DA basis we have similar double-adiabatic states which evolve according to
\bse
\be
\Phi_-  (\alpha_2)=\ket{b_{g_2}} \cos   (\alpha_2)-\ket{b_{e_2}} \sin   (\alpha_2),
\ee
\be
\Phi_+  (\alpha_2)=\ket{b_{g_2}} \sin   (\alpha_2)+\ket{b_{e_2}} \cos   (\alpha_2),
\ee
\ese
depending on the double-adiabatic angle.
Changing $\alpha_2,$ such that we realise CPT we can transfer one set of superposition states to another. For example a change from $0$ to $\pi/2$ generates,
\be\label{W-superpositions-CPT}
\ket{b_{g_2}} \stackrel{0\from \alpha_2}{\longleftarrow } \Phi_-  (\alpha_2) \stackrel{\alpha_2 \to \pi/2}{\longrightarrow } - \ket{b_{e_2}} .
\ee
\begin{figure}[tb]
\bt{cc}
\qquad \quad \includegraphics[width=0.72\columnwidth]{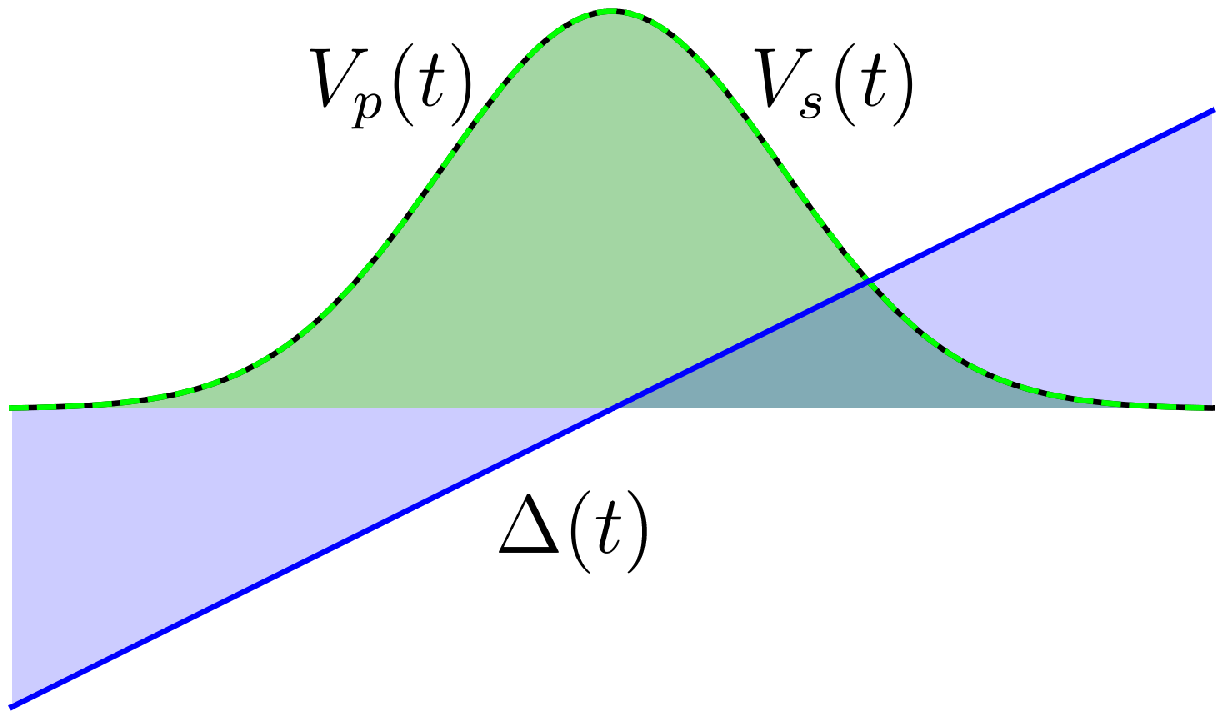}\llap{
  \parbox[b]{4.5in}{  (a)\\\rule{0ex}{1.21in}
  }}\\
\qquad \quad \includegraphics[width=0.72\columnwidth]{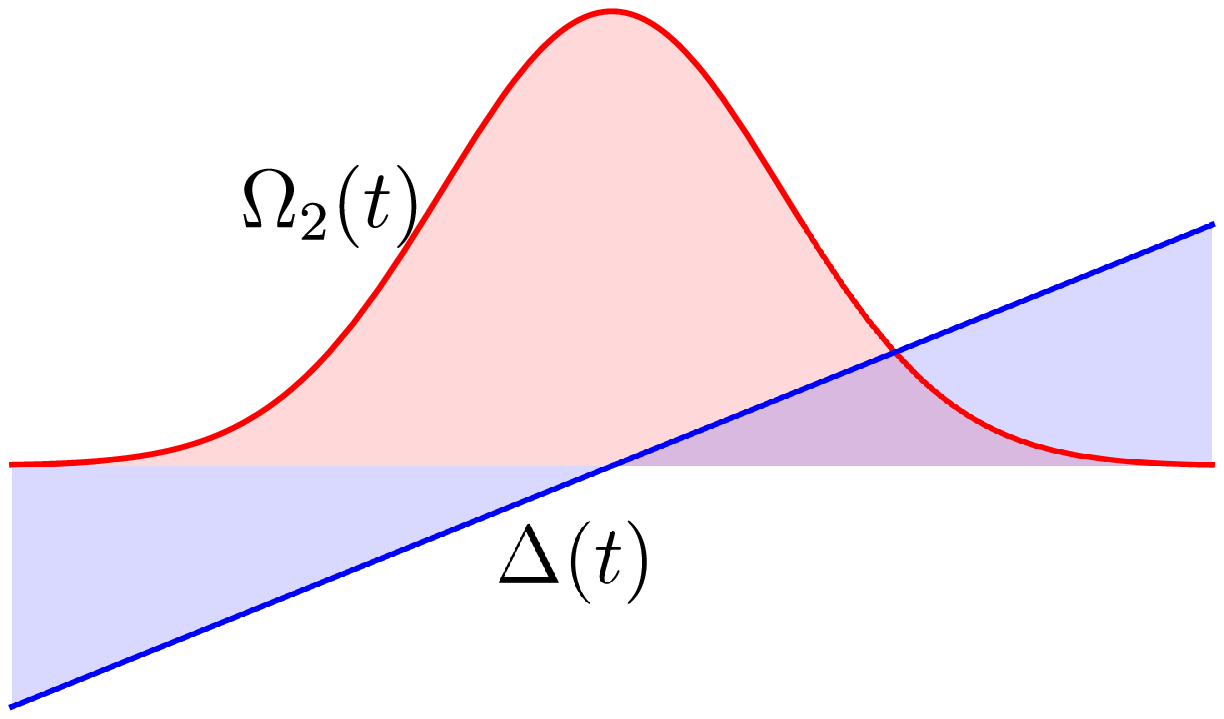}\llap{
  \parbox[b]{4.5in}{  (b)\\\rule{0ex}{1.21in}
  }}\\
\includegraphics[width=0.85\columnwidth]{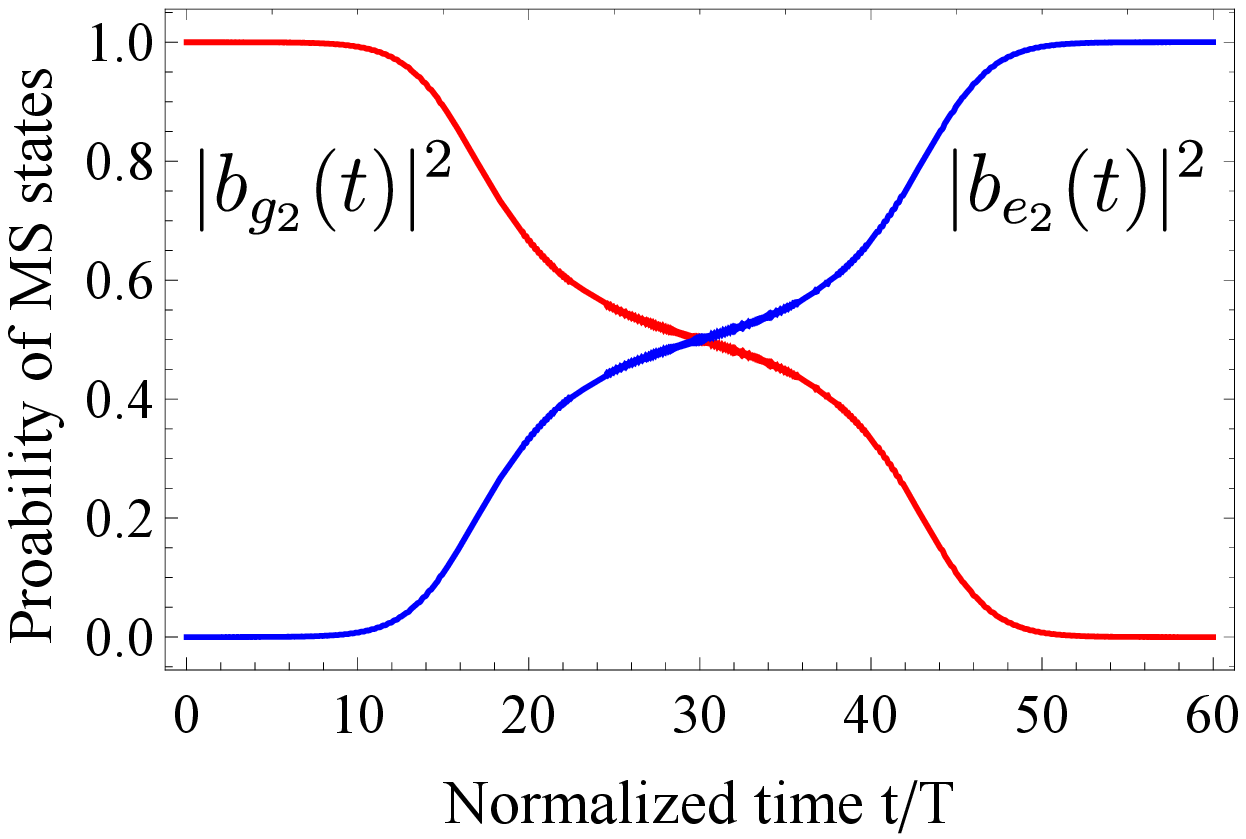}\llap{
  \parbox[b]{4.5in}{  (c)\\\rule{0ex}{0.661in}
  }} 
\et
\caption{  (Color online)
Rapid adiabatic passage  (RAP) population transfer between the MS states. 
  (a) The original couplings coinciding pulse shapes and the detuning.
  (b) The MS Rabi frequency and the detuning. 
  (c) Evolution of the bright states according to Eq.  (\ref{W-superpositions-CPT}). The detuning is a linear function $\Delta  (t)=\beta  (t-\tau) T^{-1}$ with $\beta=1$  and we used Gaussian pulses of the form $V_{i}\t=V_i^0\exp  (-  (t-\tau_i)^2/T^2)$ with coupling strengths $V_{s}^0=V_{p}^0=40 T^{-1}$ and time shift $\tau=30 T^{-1}.$  
}\label{fig:3}
\end{figure}
In terms of the original basis states the final superposition reads,
\be
\Phi_-  (\pi/2)=-\frac{\Big[\ket{e_1}+\ket{e_2}\Big] V_p  (t)+\Big[\ket{e_2}+\ket{e_3}\Big]
   V_s  (t)}{\sqrt{2}\Omega_2  (t)}.\label{W-DA state time control}
\ee

In addition to the control we can exert by changing $\alpha_2$ we have another control parameter, which is the ratio between the original couplings $V_p  (t)$ and $V_s  (t),$ as well as their value in the final moment of the excitation. For example if one of the couplings $V_p  (t)$ or $V_s  (t)$ vanish in the final moment, we can control which two of the three excited states will be in superposition. 
Alternatively if we set their magnitudes to be equal, which automatically satisfies Eq.  (\ref{MSadiabatic}), we can generate a superposition,
\be
\Phi_-  (\pi/2)_{|V_p=V_s}=-\frac{\ket{e_1}+2 \ket{e_2}+\ket{e_3}}{\sqrt{6}}\label{W-DA equal couplings},
\ee
which is insensitive to the couplings.
This is illustrated in Fig.~\ref{fig:3} for Gaussian pulses with linear detuning. 
For simplicity we have set all couplings to be equal, which means that when CPT occurs we have created the superposition state of Eq.  (\ref{W-DA equal couplings}). 

\emph{Superposition mixing by half-RAP.} If we  stop the double-adiabatic evolution earlier, we can generate a superposition between the MS states, which in the original basis results in a superposition between all of the states in the system. Letting $\alpha_2$ go from $0$ to $\pi/4,$ evolves the state vector according to
\be\label{W-superpositions-full}
\ket{b_{g_2}} \stackrel{0\from \alpha_2}{\longleftarrow } \Phi_-  (\alpha_2) \stackrel{\alpha_2 \to \pi/4}{\longrightarrow } \frac{\ket{b_{g_2}}}{\sqrt{2}}-\frac{\ket{b_{e_2}}}{\sqrt{2}} ,
\ee
which leaves the final superposition among the original basis states to be
\bea
\Phi_-  (\pi/2)=&-\frac{\Big[\ket{e_1}+\ket{e_2}\Big]
   V_p  (t)+\Big[\ket{e_2}+\ket{e_3}\Big] V_s  (t)}{2 \Omega _2  (t)}\nonumber\\
   &+\frac{ \ket{g_1}+\ket{g_2}}{2}.\label{W-full-final}
\ea
The final value of the couplings $V_i  (t)$ in Eq.  (\ref{W-full-final}) gives us the freedom to choose which states will participate in the superposition. 
If we again set the couplings to be equal, the phase insensitive superposition between the states in the original basis reads,
\be
\Phi_-  (\pi/2)_{|V_p=V_s}=\frac{ \ket{g_1}+\ket{g_2}}{2}-\frac{\ket{e_1}+2 \ket{e_2}+\ket{e_3}}{2 \sqrt{3}}.
\ee

\subsection{X linkage in three levels}
The last type of linkage we investigate is two-one-two   (X linkage) type of coupling between the states in three levels, as shown in Fig.~\ref{fig:1}   (c). The Hamiltonian of the system reads
\be
\H_X=\half\left[
\begin{array}{ccccc}
 0 & 0 & V_{p_1}  (t) & 0 & 0 \\
 0 & 0 & V_{s_1}  (t) & 0 & 0 \\
 V_{p_1}  (t) & V_{s_1}  (t) & 2 \Delta   (t) & V_{p_2}  (t) & V_{s_2}  (t) \\
 0 & 0 & V_{p_2}  (t) & 0 & 0 \\
 0 & 0 & V_{s_2}  (t) & 0 & 0 \\
\end{array}
\right].
\ee
In this example Eq.  (\ref{MSadiabatic}) holds if both sets of couplings $V_{i,j}  (t),$ where $i=p,s$ and $j=1,2$ satisfy the same inequality as Ineq.  (\ref{adcond3lvl}). Due to adiabatic excitation, the MS Hamiltonian is decomposed to two dark states and a single three level system,
\be
\H_X^{MS}=\left[
\begin{array}{cc}
 \H_d & \mathbf{0} \\
 \mathbf{0} & \H_b \\
\end{array}
\right],
\ee
where the bright Hamiltonian is given by Eq.  (\ref{HMS3lvl}) with coupling elements,
\bse
\be
\Omega_p  (t)=\sqrt{V_{p_1}  (t){}^2+V_{s_1}  (t){}^2},
\ee
and
\be
\Omega_s  (t)= \sqrt{V_{p_2}  (t){}^2+V_{s_2}  (t){}^2}.
\ee
\ese
In terms of the original basis states the MS state vector reads,
\be
\C_X^{MS}=\left[
\begin{array}{c}
 \frac{\ket{g_2} V_{p_1}  (t)-\ket{g_1} V_{s_1}  (t)}{\Omega _p  (t)} \\
 \frac{\ket{e_2} V_{p_2}  (t)-\ket{e_1} V_{s_2}  (t)}{\Omega _s  (t)} \\
 \frac{\ket{g_1} V_{p_1}  (t)+\ket{g_2} V_{s_1}  (t)}{\Omega _p  (t)} \\
 \ket{m} \\
 \frac{\ket{e_1} V_{p_2}  (t)+\ket{e_2} V_{s_2}  (t)}{\Omega _s  (t)} \\
\end{array}
\right],
\ee
where the first two states are dark, and the last three are bright.
We now turn to the SA basis where the state vector reads,
\be
\C^{DA}  (t)=\U  (t)\C^{MS}=\left[
\begin{array}{ccc}
 1 & 0 & 0 \\
 0 & 1 & 0 \\
 0 & 0 & \R\left  (\theta,\phi\right) \\
\end{array}
\right]\left[\begin{array}{c}
 \ket{d_1} \\
 \ket{d_2} \\
 \ket{\B}\\
\end{array}\right],\label{X-DA states}
\ee
with the rotation matrix now given by Eq.   (\ref{Rot3lvl}), and we again assume that the adiabatic  (Eq.  (\ref{MSadiabatic})) and the double-adiabatic conditions  (Eq.   (\ref{adcond3lvl})) hold for the couplings and the detuning. The type of superpositions we can generate fall under the following categories: \\
\emph{Superposition transfer by STIRAP.} The evolution of the DA states in terms of the MS states reads,
\bse
\bea
&\Phi_+  (\theta,\phi)= \sin   (\phi ) \Big[ \ket{b_1} \sin   (\theta )+\ket{b_2} \cos   (\theta )\Big]+\ket{m} \cos   (\phi ), \\
&\Phi_-  (\theta,\phi)= \cos   (\phi ) \Big[ \ket{b_1} \sin   (\theta )+\ket{b_2} \cos   (\theta )\Big]-\ket{m} \sin   (\phi ),\\
&\Phi_0  (\theta)=\ket{b_1} \cos   (\theta )-\ket{b_2} \sin   (\theta ).
\ea
\ese
We are interested in $\Phi_0  (\theta)$ specifically, since it does not involve the intermediate state $\ket{m}.$ In order to transfer the population from the ground to the excited MS state we employ STIRAP mechanism. This means a counter intuitive pulse order, that is $\Omega_s  (t)$ before $\Omega_p  (t).$ In terms of our original coupling elements this requires that the excitation between the final and intermediate levels has to start before the excitation of the ground and intermediate levels.
Rotation of the double-adiabatic angle from $0$ to $\pi/2$ results in
\be\label{X-superpositions-full}
\ket{b_{g}} \stackrel{0\from \theta}{\longleftarrow } \Phi_0  (\theta) \stackrel{\theta \to \pi/2}{\longrightarrow } -\ket{b_{e}}.
\ee
In terms of the original states the final superposition is given by
\be
\Phi_0  (\pi/2)=-\frac{\ket{e_1} V_{p_2}  (t)+\ket{e_2} V_{s_2}  (t)}{ \Omega _s  (t){}^2}.
\ee
In a similar way to the W-linkage, the final superposition also depends on the final values of the couplings. However, here we have the additional requirement that $\Omega_s  (t)/\Omega_p  (t)\to0$ in the end of the interaction, in order to make $\Phi_0  (\pi/2)$  coincide with $\ket{b_e}.$ This means that the only way we can retain the superposition is if we set $V_{p_2}  (t)=V_{s_2}  (t).$ In this case the final superposition reads,
\be
\Phi_0  (\pi/2)_{|V_{p_2}  (t)=V_{s_2}  (t)}=-\frac{\ket{e_1} +\ket{e_2} }{ \sqrt{2}}.
\ee
With the above procedure we can only transfer equal superposition states, which is useful if we want the target superposition to be in the same level.\\
\emph{Superposition mixing by fractional STIRAP.} In a similar way to the example for W system, if we stop the DA angle at $\theta=\pi/4,$ which is known as fractional STIRAP, we can generate a superposition among the MS states,
\bea
\ket{b_{g}} \stackrel{0\from \theta}{\longleftarrow } \Phi_0  (\theta) \stackrel{\theta \to \pi/4}{\longrightarrow } \frac{\ket{b_{g}}}{\sqrt{2}}-\frac{\ket{b_{e}}}{\sqrt{2}}.\label{X-f STIRAP}
\ea
The requirement for f-STIRAP is to have the pump and the Stokes pulses vanish simultaneously. 
\begin{figure}[tb]
\bt{cc}
\qquad \quad \includegraphics[width=0.74\columnwidth]{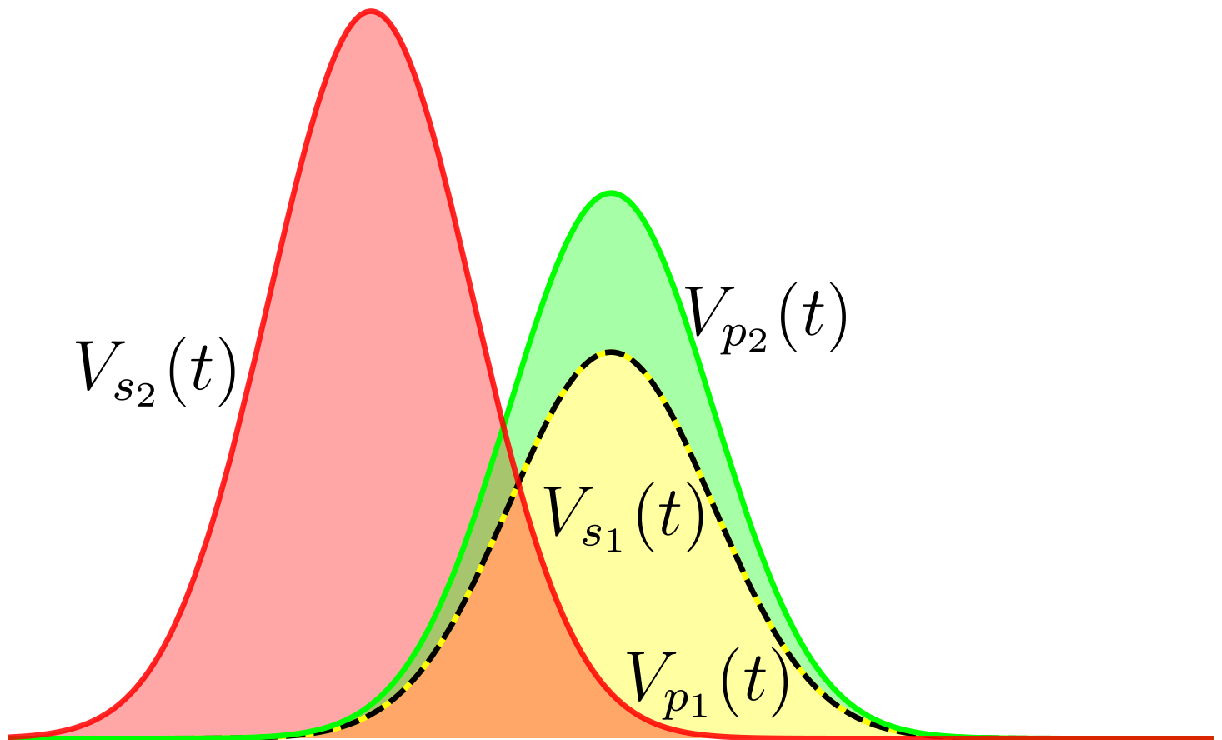}\llap{
  \parbox[b]{4.5in}{  (a)\\\rule{0ex}{1.31in}
  }}\vspace{.2cm}\\
 \qquad \quad \includegraphics[width=0.74\columnwidth]{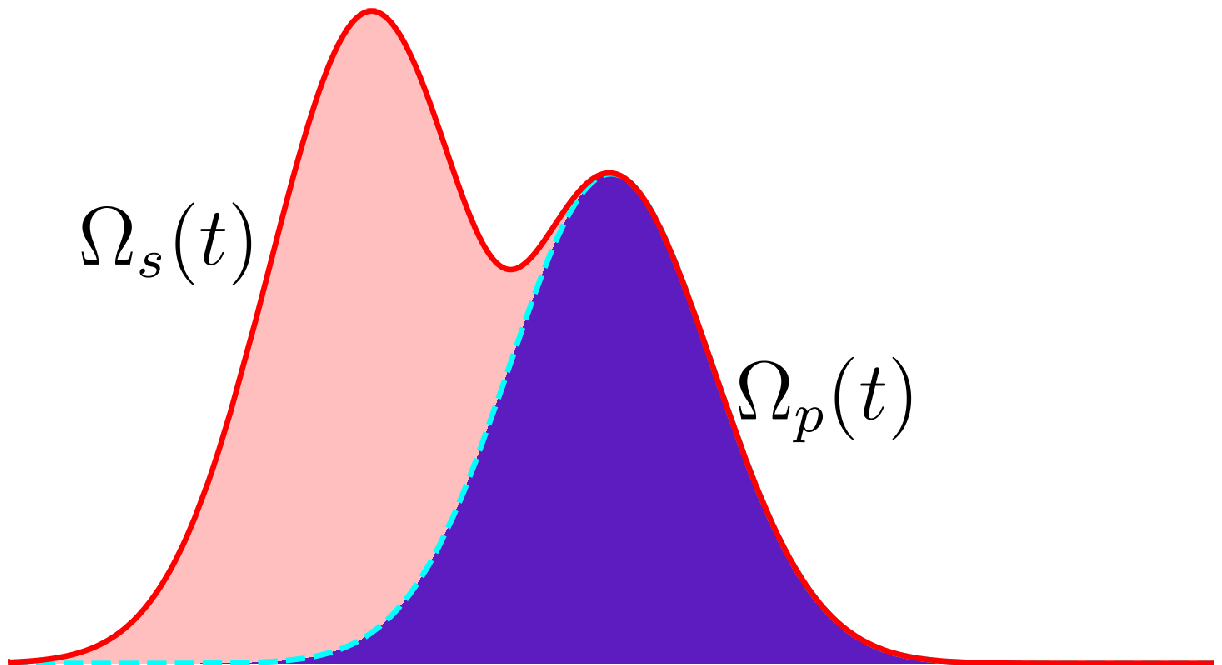}\llap{
  \parbox[b]{4.5in}{  (b)\\\rule{0ex}{1.261in}
  }}\vspace{.2cm}\\
  \includegraphics[width=0.9\columnwidth]{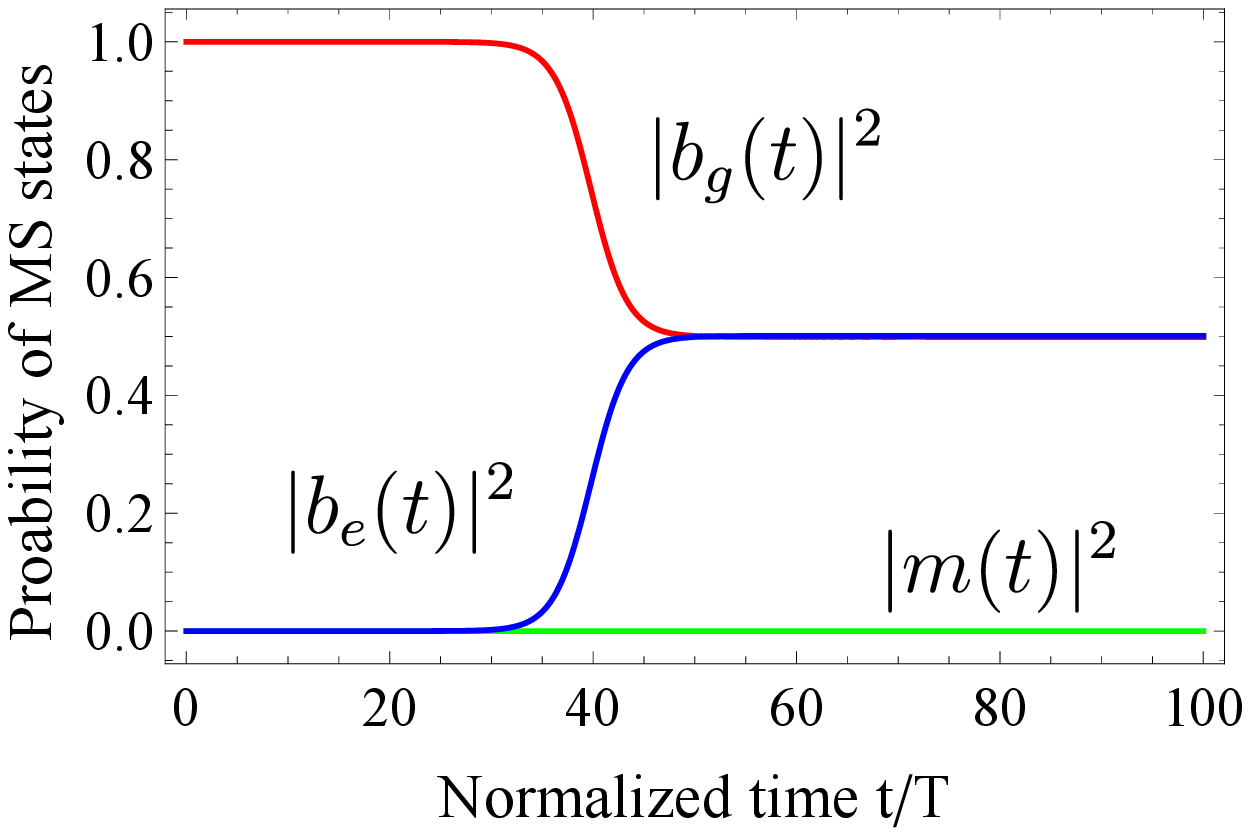}\llap{
  \parbox[b]{4.6in}{  (c)\\\rule{0ex}{1.71in}
  }}  
\et
\caption{  (Color online)
Fractional STIRAP population transfer between the MS states.
  (a) The original couplings in counter intuitive arrangement, where $V_{s_2}  (t)$  start before $V_{p_2}  (t)$ and $V_{p_1}  (t)=V_{s_1}  (t)$.
 Consequently in   (b) the MS Rabi frequency $\Omega_s  (t)$ comes before $\Omega_p  (t)$, but they vanish simultaneously.
The   (c) frame demonstrates the evolution of the bright states according to Eq.  (\ref{X-f STIRAP}). Since  $V_{s_2}  (t)$ ends before $V_{p_2}  (t)$ the final superposition is the one of Eq.  (\ref{X-3state superposition}). The simulation is under constant detuning $\Delta=0.0001 T^{-1},$ and Gaussian pulses of the form $V_{i}\t=V_i^0\exp  (-  (t-\tau_i)^2/T^2)$ with strength of the couplings $V_{s2}^0=160 T^{-1},V_{p2}^0=120 T^{-1},V_{s1}^0=V_{p1}^0=85 T^{-1}$ and the pulses are shifted by $\tau_{s_2}=30 T^{-1},$ and $\tau_{p_2}=\tau_{s_1}=\tau_{p_1}=50 T^{-1}.$  
}\label{fig:4}
\end{figure}
 In our current system this translates to the generation of
\bea
\Phi_0  (\pi/4)_{|V_{p_1}  (t)=V_{s_1}  (t)}=&-\frac{\sqrt{2} \Big[\ket{e_1} V_{p_2}  (t)+\ket{e_2}
   V_{s_2}  (t)\Big]}{2\Omega _s  (t)} \nonumber \\&+\frac{\ket{g_1}+\ket{g_2}}{2},
\ea
where for simplicity we have set $V_{p_1}  (t)=V_{s_1}  (t).$
Unlike the case of the standard STIRAP population transfer, here we can still control the final superposition state by the values of the $V_{i_2}  (t)$ couplings. If either of them vanishes before the rest we can generate a superposition of three instead of four states. Such example is shown in Fig.~\ref{fig:4}, where $V_{s_2}  (t)$ vanishes before all other couplings and thus the final superposition reads
\be
\Phi_0  (\pi/4)_{|V_{p_1}  (t)=V_{s_1}  (t)}=\frac{\ket{g_1}+\ket{g_2}}{2}-\frac{\sqrt{2} \ket{e_1} }{2}.\label{X-3state superposition}
\ee
 This procedure is quite flexible and is a great choice whenever we want to generate a superposition between states from different levels, since it allows a control over the participant states by the pulse delays. 


\section{Conclusions}\label{sec:four}


In this paper we have described a procedure for creating coherent superpositions of $N$ quantum states distributed over two or three levels.
We explored the evolution of the system under adiabatic, and further double-adiabatic conditions which allowed us to apply the time dependent MS transformation
and eliminate the non-adiabatic couplings such that the final state of the system depends on the alignment of the double-adiabatic vector with the desired superposition state. The adiabatic/double-adiabatic evolution also ensured that any imperfections in the pulse shapes will not affect the target state, since it depends only on their asymptotic values.
We investigated three types of systems, namely the N-pod, W and X-linkages. In each we showed how specific superposition states can be generated by adiabatic techniques. In the N-pod system we showed how the system can be initialized in a bright state without populating the dark states, by letting one of the pulses to excite the system before the rest. Then we showed how a specific superposition can be generated by an excitation that achieves CPR and the participant states can be picked by switching off prematurely the  couplings corresponding to the unwanted states. This type of excitation allows us to generate a qubit state within a degenerate multistate level without the need to isolate the two states from the rest of the system.
In the W system we showed how coinciding pulses with asymmetric detuning can achieve CPT which resulted in superposition transfer from the ground level to the excited level. In addition we showed how a superposition among all of the states in the system can be generated by a half-RAP excitation. In the X system, we demonstrated how STIRAP can transfer the superposition from the ground level to the excited level without populating the intermediate level. Alternatively we demonstrated how a superposition between the ground and the excited states can be generated via fractional-STIRAP. The later technique gave us the ability to eliminate one of the excited states in the superposition.
This is in contrast to the previous work of Shapiro et al. \cite{Shapiro,Amitay,Thanopulos,Kral}, where there are only two different time-dependent shapes for the Rabi frequencies, pump and Stokes shapes, and if the Rabi frequencies obey one of these forms, there is a special dark state like in STIRAP \cite{Vitanov2001A,Vitanov2001B} that is not connected to the other dark states and is a superposition of all ground states.
The presented procedure has a significant potential for manipulation of multistate quantum bits in quantum information processing, for example, in designing arbitrary unitary gates and also paves the way towards qudit quantum gates.
\section*{Acknowledgements}
This paper is dedicated to the memory of our dear friend Bruce W. Shore, who inspired and encouraged us throughout the years to do what we love and love what we do.
This work is supported by the European Commission's Horizon-2020 Flagship on
Quantum Technologies project 820314   (MicroQC). 
KNZ acknowledges support by the Bulgarian NSF on projcet MSPLICS with Grant number KP-06-DB/4.


\end{document}